\documentclass[prd,aps,superscriptaddress,twocolumn,%
showpacs,preprintnumbers,amsmath,amssymb,nofootinbib]{revtex4-2}

\usepackage{graphicx}
\usepackage{bm}
\usepackage[colorlinks=true, pdfstartview=FitV, citecolor=blue, urlcolor=blue]{hyperref}
\usepackage{slashed}
\usepackage{array}
\usepackage{mathtools}

\graphicspath{{./Figures/}}

\newcommand{\diff}{\mathrm{d}}
\newcommand{\p}{\partial}

\newcommand{\be}{\begin{equation}}      
\newcommand{\ee}{\end{equation}}      
\newcommand{\bea}{\begin{eqnarray}}      
\newcommand{\eea}{\end{eqnarray}}

\newcommand{\tr}{\mathrm{tr}}
\newcommand{\im}{\mathrm{i}}

\newcommand{\calA}{\mathcal{A}}

\newcommand{\rme}{\mathrm{e}}

\newcommand{\bra}{\big\langle}
\newcommand{\ket}{\big\rangle}

\begin{document}

\preprint{YITP-21-29}

\title{Non-invertible 1-form symmetry and Casimir scaling in 2d Yang--Mills theory}

\author{Mendel Nguyen}
\email{mendelnguyen@gmail.com} 
\affiliation{Department of Physics, North Carolina State University, Raleigh, NC 27607, USA}

\author{Yuya Tanizaki}
\email{yuya.tanizaki@yukawa.kyoto-u.ac.jp}
\affiliation{Yukawa Institute for Theoretical Physics, Kyoto University, 
  Kyoto 606-8502, Japan}

\author{Mithat \"Unsal}
\email{unsal.mithat@gmail.com}
\affiliation{Department of Physics, North Carolina State University, Raleigh, NC 27607, USA}

\begin{abstract}
Pure Yang--Mills theory in $2$ spacetime dimensions shows exact Casimir scaling. Thus there are infinitely many string tensions, 
and this has been understood as a result of non-propagating gluons in $2$ dimensions. From ordinary symmetry considerations, however, this richness in the spectrum of string tensions seems mysterious. 
Conventional wisdom has it that it is the center symmetry that classifies string tensions, but being finite it cannot explain infinitely many confining strings. 
In this note, we resolve this discrepancy between dynamics and kinematics by pointing out the existence of a non-invertible 1-form symmetry, which is able to distinguish Wilson loops in different representations. 
We speculate on possible implications for Yang--Mills theories in 3 and 4 dimensions.
\end{abstract}

\maketitle
\section{Introduction}
\label{sec:introduction}

\begin{figure}[t]
\centering
\vspace{-0.2cm}
\includegraphics[width = 0.45 \textwidth]{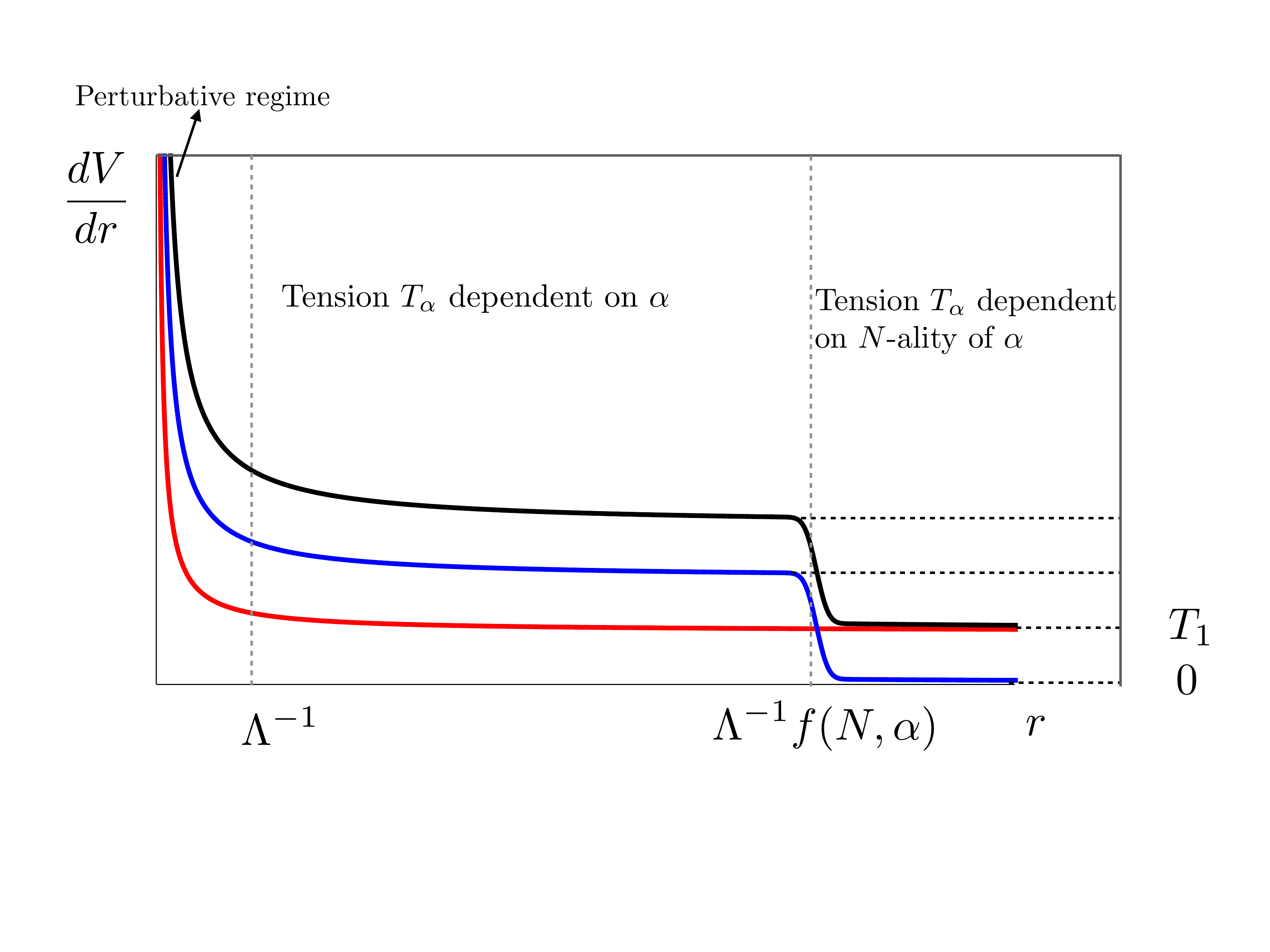}
\vspace{-1.7cm}
\caption{Characteristic behavior of derivative of inter-quark potential in some representations $\alpha$ 
in $d=3,4$ $\mathrm{SU}(N)$ Yang--Mills theory. 
Here, we take Young tableaux  $\alpha=(1,0,\ldots, 0), (1,0,\ldots,0,1), (2,0,\ldots,0,1)$ for red, blue, and black curves, respectively. 
 }
\label{fig:tension}
\vspace{-0.5cm}
\end{figure}

In quantum gauge theories, string tensions are characteristic properties of confinement phenomena, as they specify the static quark-antiquark potentials. 
We can compute them theoretically as expectation values of Wilson loops for the gauge representation $\alpha$ of the test quark. 
For confining $\mathrm{SU}(N)$ Yang--Mills theory in $3$ or $4$ dimensions, it is expected that Wilson loops behave very differently over three different length scales (see Fig.~\ref{fig:tension})~\cite{Ambjorn:1984dp, Poulis:1995nn, Bali:2000un, Philipsen:1999wf, Stephenson:1999kh, deForcrand:1999kr, Greensite:2003bk, Wellegehausen:2010ai}: 
\begin{enumerate}
    \item At short distances, the potential obeys Coulomb's law, with a coefficient given by the Casimir invariant $c_{\alpha}$.
    \item At intermediate distances, the potential becomes linear, with a string tension $T_{\alpha}$ depending on $\alpha$. 
    \item At long distances, the potential remains linear, but the string tension $T_{\alpha}$ depends only on $N$-ality of $\alpha$.
\end{enumerate} 
The last property can be understood as a result of string breaking via soft-gluon exchange. 
Moreover, we can nicely describe the relevant selection rule using the center symmetry~\cite{tHooft:1977nqb, Polyakov:1978vu}, or the $1$-form symmetry~\cite{Gaiotto:2014kfa, Sharpe:2015mja}. 
However, this is not the whole story of confinement. 
In particular, the behavior at intermediate distances is curious: the theory is already confining, but the string tensions are not characterized by center symmetry alone.

Driven by this curiosity, the authors were led in Ref.~\cite{Nguyen:2021yld} to explore a similar phenomenon in a simpler confining gauge theory in 3 dimensions, where string tensions at any distance scale do not obey the $N$-ality rule. 
In that investigation, it was realized that a \emph{non-invertible} 1-form symmetry is present in that model, and that it can distinguish Wilson loops in different representations even when their $N$-alities are the same. 
Thus, the following question naturally arises: Does non-invertible $1$-form symmetry also exist in Yang--Mills theory? 
If so, can it be used to classify various string tensions at intermediate distances? 

In this note, we consider properties of the confining strings of pure Yang--Mills theory in 2 spacetime dimensions. It is well-known that 2d Yang--Mills theory is exactly solvable~\cite{Migdal:1975zg, Witten:1991we}, and the string tensions obey the Casimir scaling law, 
which says that the confining force is proportional to the Casimir invariant $c_{\alpha}$ of the test quark. 
In this case, we can understand from  dynamical considerations why the string tensions need not be characterized by center symmetry. 
In $2$ dimensions, there are no propagating gluons, and string breaking never occurs.
Nevertheless, it behooves us to explain this phenomenon purely from considerations of symmetry, which must be an important step towards the harder cases of Yang--Mills theories in higher dimensions.

We find that there is in fact a good symmetry-based justification for the rich spectrum of confining strings in 2d Yang--Mills theory. 
After a brief review of the exact solution, we define a topological point-like disorder operator that can distinguish Wilson loops in different representations of the gauge group. At the end, we speculate on possible implications for the behavior of the confining strings of Yang--Mills theories in 3 and 4 dimensions at intermediate distances.


\section{2d Yang--Mills and Casimir scaling}
\label{sec:review-2dYM}

Let us begin by reviewing some exact results in 2d Yang--Mills theory~\cite{Migdal:1975zg, Witten:1991we}.
Let $G$ be an arbitrary gauge group, which is assumed to be simple, connected, and simply-connected.

Pure Yang--Mills theory on a spacetime $X$ with Riemannian metric $\diff s^2=g_{ij}\diff x^i \otimes \diff x^j$ is described by the action
\begin{equation}
    S
    = - \frac{1}{e^2} \int_X \tr(F \wedge *F),
    \label{YM-action}
\end{equation}
where $F$ is the field strength of the $G$--gauge field $A$,
\begin{equation}
    F = \diff A + A \wedge A. 
\end{equation}
A particularly special feature of the theory in 2 dimensions is its invariance under area-preserving diffeomorphisms. 
This becomes self-evident when we rewrite the action \eqref{YM-action} in terms of the adjoint scalar $\phi = *F$:
\begin{equation}
    S = - \frac{1}{e^2} \int_X \tr(\phi \wedge *\phi) 
    = - \frac{1}{e^2} \int_X \tr(\phi^2) \sqrt{g}\, \diff^2 x, 
\end{equation}
where $g=\det(g_{ij})$. Here, it should be noted that the metric $g_{ij}$ enters only through the area form $\sqrt{g}\diff^2 x$. 

As the only invariant of a top-degree form is its total integral, the partition function can depend on the metric $g_{ij}$ only through the total area $\calA = \int_X \sqrt{g} \diff^2 x$. Furthermore, since in 2 dimensions the gauge coupling $e$ has the dimensions of inverse-length, the partition function is a function of the dimensionless combination $e^2 \cal A$. In view of these considerations, we set $e=1$ and denote the partition function by $Z_X(\calA)$.

For our purposes, it is most useful to work with a particular lattice regularization known as the heat-kernel formulation, or the generalized Villain formulation~\cite{Migdal:1975zg, Drouffe:1978py, Menotti:1981ry, Lang:1981rj}. Let us denote links by $\ell$, $G$-valued link-variables by $U_\ell$, plaquettes by $p$, and holonomies around $\p p$ by $U_p \coloneqq \mathcal{P}\prod_{\ell\in \p p}U_\ell$, where $\mathcal{P}$ denotes path ordering. 
Here, there is no restriction on the shapes of the plaquettes; they can be any polygons (e.g., triangles, squares, pentagons, etc.).
The heat-kernel lattice formulation is defined by taking the one-plaquette weight to be
\begin{equation}
    Z_\triangle (U_p, \calA_p) 
    = \sum_{\alpha} d_\alpha \chi_\alpha (U_p) \exp\left(-c_\alpha \calA_p\right),
    \label{eq:weight_heat_kernel}
\end{equation}
where $\alpha$ runs through unitary irreducible representations of $G$, $c_\alpha$ denotes the quadratic Casimir invariant of $\alpha$,
$\chi_\alpha:G\to \mathbb{C}$ is the character of $\alpha$, $d_\alpha=\chi_\alpha(\mathbb{I})$ is the dimension of $\alpha$, and $\calA_p$ denotes the area of $p$. 
Here, $\mathbb{I}\in G$ is the identity element. 
The partition function is then given by
\begin{equation}
    Z_X (\calA)
    = \int \prod_\ell \diff U_\ell \prod_p Z_\triangle (U_p,\calA_p), 
    \label{eq:pf_heat_kernel}
\end{equation}
where the link-variables are integrated with respect to the normalized Haar measure of $G$.


In $2$ spacetime dimensions, the heat-kernel formulation has a remarkable property: the partition function~(\ref{eq:pf_heat_kernel}) is invariant under subdivisions. 
More precisely, if two plaquettes $p_1,p_2$ meet at a common link $\ell$, then one has the `sewing' property:
\begin{multline}
    \int \diff U_\ell Z_{\triangle}(U_{p_1},\calA_{p_1})Z_{\triangle}(U_{p_2},\calA_{p_2}) \\
    = Z_{\triangle}(U_{p_1\cup p_2}, \calA_{p_1} + \calA_{p_2}).
    \label{eq:sewing}
\end{multline}
To see this, 
let us write the one-plaquette weights on the left-hand-side explicitly as
\begin{align}
    Z_\triangle (U_{p_1}, \calA_{p_1}) 
    &= \sum_{\alpha} d_\alpha \chi_\alpha (U_1 U_\ell) \exp\left(-c_\alpha \calA_{p_1} \right), \\
    Z_\triangle (U_{p_2}, \calA_{p_2})
    &= \sum_{\beta} d_\beta \chi_\beta (U^{-1}_{\ell} U_2) \exp\left(-c_\alpha \calA_{p_2} \right), 
\end{align}
where we set $U_{p_1}=U_1 U_\ell$, $U_{p_2}=U_\ell^{-1}U_2$ so that $U_{p_1 \cup p_2}=U_{1}U_2$. 
Using a formula on characters,
\begin{equation}
    \int \diff g\, \chi_\alpha (a g) \chi_\beta (g^{-1} b) 
    = \delta_{\alpha,\beta} \frac{\chi_{\alpha}(ab)}{d_{\alpha}},
    \label{eq:character-1}
\end{equation}
we readily obtain Eq.~(\ref{eq:sewing}). 
Therefore, this specific lattice formulation is already at the fixed point of the renormalization group, and reproduces the results of the continuum theory.

The partition function on any genus-$g$ spacetime $X=\Sigma_g$ is readily obtained as
\begin{equation}
    Z_{\Sigma_g}(\calA)
    = \sum_\alpha d_\alpha^{2-2g} \exp\left(-c_\alpha \calA\right).
\end{equation}
We note that, in the infinite area limit, which will give us the partition function on $\mathbb{R}^2$, we have
\begin{equation}
    Z_{\mathbb{R}^2}
    = Z_{\Sigma_g}(\calA \to \infty) = 1,
\end{equation}
as only the trivial representation $\alpha = \bm 1$ contributes. 

\begin{figure}[t]
\begin{center}
\includegraphics[width = 0.26 \textwidth]{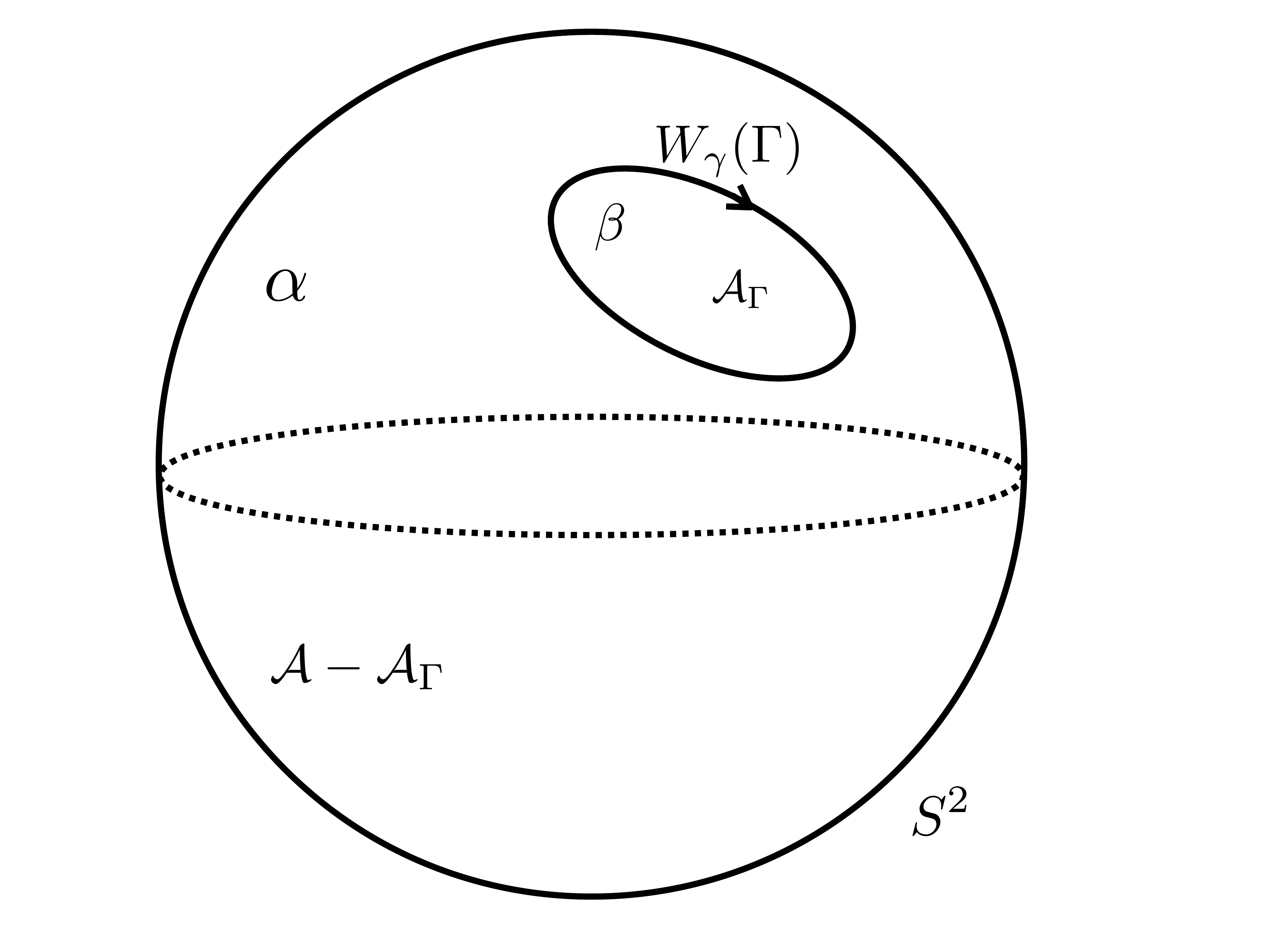}
\caption{Expectation value of a Wilson loop $W_\gamma(\Gamma)$ on $S^2$.  }
\label{fig:Wilson}
\vspace{-0.8cm}
\end{center}
\end{figure}

Let us now compute the expectation value of a single Wilson loop $W_\gamma(\Gamma)$ on $\mathbb{R}^2$. 
For convenience, we initially take $X=S^2$, and then take $\calA \to \infty$ at the end while keeping the area  $\calA_\Gamma$ `enclosed' by $\Gamma$ held fixed (see Fig.~\ref{fig:Wilson}). 
We have
\begin{align}
    \bra W_\gamma(\Gamma) \ket_{S^2}
    &= \frac{1}{Z_{S^2}} \sum_{\alpha, \beta} d_\alpha d_\beta \, \rme^{-c_\alpha(\calA - \calA_\Gamma) - c_\beta \calA_\Gamma} \nonumber\\
    & \quad \times \int \diff U \chi_\alpha(U) \chi_\gamma(U) \chi_\beta(U^{-1}).
\end{align}
We then use another formula on characters,
\begin{equation}
    \int \diff g\, \chi_\alpha(g) \chi_\gamma(g) \chi_\beta(g^{-1})
    = N^{\beta}_{\alpha \gamma},
    \label{eq:character-2}
\end{equation}
where $N^{\beta}_{\alpha \gamma}$ is the multiplicity of $\beta$ in the decomposition of $\alpha \otimes \gamma$ into irreducible representations. This gives the Wilson loop average on $S^2$ as
\begin{equation}
    \bra W_\gamma(\Gamma) \ket_{S^2}
    = \frac{1}{Z_{S^2}} \sum_{\alpha, \beta} N^{\beta}_{\alpha \gamma} d_\alpha d_\beta \,  \rme^{-c_\alpha(\calA - \calA_\Gamma) - c_\beta \calA_\Gamma} .
\end{equation}
Taking the infinite area limit, only $\alpha = \bm 1$ contributes as above, and as we clearly have $N^{\beta}_{ \bm 1 \gamma} = \delta_{\beta,\gamma}$, it follows that the Wilson loop average on $\mathbb{R}^2$ is given by
\begin{equation}
    \bra W_\gamma(\Gamma) \ket_{\mathbb{R}^2}
    = d_\gamma \exp\left(-c_\gamma \calA_\Gamma\right). 
    \label{wilson-vev}
\end{equation}
Thus, Wilson loops in all representations obey area law decay, and the string tensions are precisely dictated by Casimir scaling:
\begin{equation}
    T_{\gamma} = c_\gamma.
\end{equation}
In particular, $2$d pure Yang--Mills theory has infinitely many string tensions, which cannot be solely characterized by the center symmetry, or `$N$-ality'. 
Is there any kinematical way to understand this result? 


\section{Non-invertible 1-form symmetry}

For a spacetime point $x$ and a conjugacy class $[U]=\{gUg^{-1}\,|\, g\in G\}$ of $G$, we define a disorder operator $V_{[U]}(x)$ by the following injunction: \begin{itemize}
    \item Delete the point $x$ from spacetime, and perform the path integral over gauge fields $A$ with holonomy $\mathrm{hol}_{C}(A) \in [U]$ for small clockwise-oriented circles $C$ surrounding $x$.
\end{itemize}
This is a $2$d version of Gukov--Witten surface operators in $4$d gauge theories~\cite{Gukov:2008sn, Gukov:2014gja}. 
We note that this operator is gauge-invariant, as we fix the conjugacy class of the holonomy instead of the holonomy itself. 
Furthermore, this operator is topological thanks to the invariance under area-preserving diffeomorphisms. 
But as we will see by explicit calculation, this operator does not necessarily have an inverse. 
Hence, we must view $V_{[U]}(x)$ as the generator of a \emph{non-invertible 1-form symmetry}. 

From a modern perspective on symmetry in relativistic field theory, a conservation law is interpreted as the existence of topological operators~\cite{Gaiotto:2014kfa}. 
Non-invertible symmetry is a new kind of symmetry based on this idea, but the requirement that the symmetry elements obey a group-like multiplication law is relaxed. 
So far, the utility of this notion has been demonstrated mainly in the context of $2$d field theories~\cite{Bhardwaj:2017xup, Buican:2017rxc, Freed:2018cec,  Chang:2018iay,Thorngren:2019iar,Ji:2019jhk, Komargodski:2020mxz, Aasen:2020jwb}.\footnote{For applications to $3$d gauge theories, see Refs.~\cite{Rudelius:2020orz, Nguyen:2021yld}. } 
Our operator $V_{[U]}(x)$ has an analogous property, but it acts on line operators instead of point-like operators. 

Using the heat-kernel lattice formulation, we can easily compute correlation functions of these disorder operators. All one needs to do is to pick out an infinitesimal plaquette $p$ containing the point $x$ of the dual lattice and a representative $U$ of the conjugacy class $[U]$, and then fix the path-ordered product of link-variables $U_p$ to be $U^{-1}$. 

For example, the $n$-point function of the disorder operators on $S^2$ can be computed as\footnote{We note that in the axiomatic approach to 2d Yang--Mills theory, the 2- and 3-point functions here are precisely the `cylinder' and `pants' amplitudes from which all other amplitudes are built according to the general cutting-and-gluing law~\cite{Witten:1991we}.}
\begin{equation}
    \Bigl\langle \prod_{i=1}^{n} V_{[U_i]}(x_i) \Bigr\rangle_{S^2}
    = \frac{1}{Z_{S^2}}
    \sum_\alpha d_\alpha^2 \,  \rme^{-c_{\alpha} \calA} 
    \prod_{i=1}^{n} \frac{\chi_\alpha(U_i)}{d_\alpha}.
\end{equation}
In particular, the locations of the insertion points $x_{i}$ do not appear in the vacuum expectation values, which confirms that these operators are topological. 

For us, the important thing about the disorder operators $V_{[U]}(x)$ is that
they act non-trivially on Wilson loops. What's more, they can distinguish Wilson loops in different representations. In particular, on $\mathbb{R}^2$, we have 
\begin{equation}
    \bra W_\gamma(\Gamma) V_{[U]}(x) \ket =
    \begin{cases}\displaystyle
    \frac{\chi_\gamma(U)}{d_\gamma} \bra W_\gamma(\Gamma) \ket &  \text{for $x$ inside $\Gamma$}, \\
    \bra W_\gamma(\Gamma) \ket &  \text{for $x$ outside $\Gamma$}. 
    \end{cases}
    \label{eq:WV}
\end{equation}
Let us point out that the non-invertible 1-form symmetry generated by the $V_{[U]}(x)$ actually contains the 1-form center symmetry as a special case, which is similar to the $3$d semi-Abelian theory~\cite{Nguyen:2021yld}.  
Namely, the 1-form center symmetry is generated by the $V_{[U]}(x)$ with $U$ in the center of $G$. 
For $G = \mathrm{SU}(N)$, the center elements can be written as $U = \omega \mathbb{I}$ with $\omega^N = 1$, and we have 
\be
\frac{\chi_\gamma (\omega \mathbb{I})}{d_\gamma} = \omega^{|\gamma|},
\ee
where $|\gamma|$ is the $N$-ality of $\gamma$.
Thus, for this specific choice, $V_{[\omega \mathbb{I}]}(x)$ is invertible. 

\begin{figure}[t]
\centering
\includegraphics[width = 0.45 \textwidth]{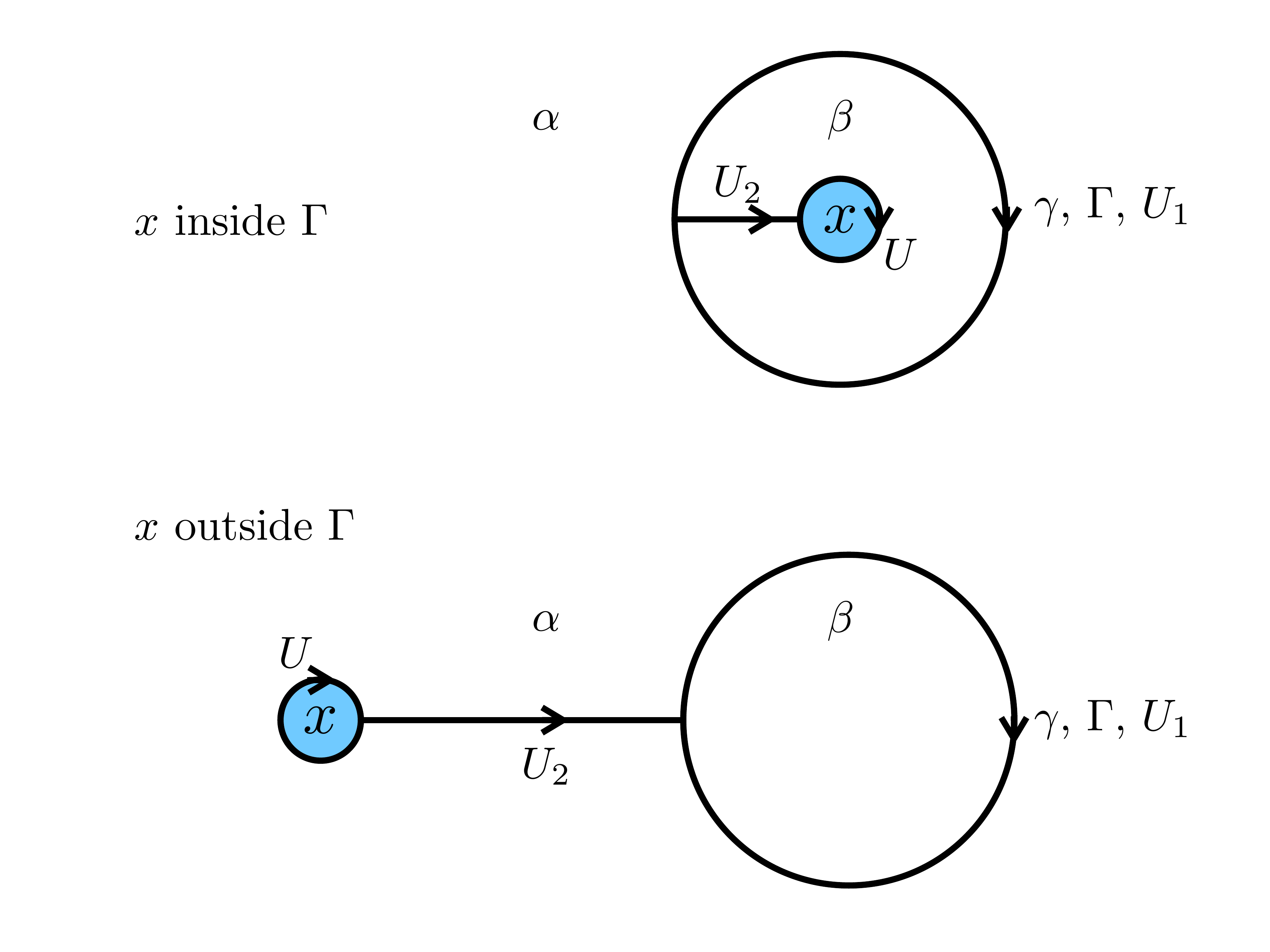}
\vspace{-0.5cm}
\caption{Defect operator $V_{[U]}(x)$ inside and outside of the Wilson loop $W_{\gamma}(\Gamma)$.}
\label{fig:defect}
\vspace{-0.5cm}
\end{figure}

Let us now prove Eq.~\eqref{eq:WV}. As before, we initially work on $S^2$ and take the infinite area limit at the end. Consider first the case where $x$ is inside $\Gamma$. According to Fig.~\ref{fig:defect}, we get
\begin{align}
    &\bra W_\gamma(\Gamma) V_{[U]}(x) \ket_{S^2} \nonumber \\ 
        &= \frac{1}{Z_{S^2}} \sum_{\alpha, \beta} d_\alpha d_\beta\,  \exp(-c_\alpha(\calA - \calA_\Gamma) - c_\beta \calA_\Gamma) \nonumber \\
        & \quad \times \int \diff U_1 \diff U_2 \chi_\alpha (U_1) \chi_\gamma (U_1) \chi_\beta (U_2 U_1^{-1} U_2^{-1} U). 
        \label{eq:WV-linked}
\end{align}
We can easily evaluate the group integrals with the help of yet another formula on characters,
\begin{equation}
    \int \diff g\, \chi_\alpha (g a g^{-1} b) = \frac{\chi_\alpha(a) \chi_\alpha(b)}{d_\alpha},
    \label{eq:character-3}
\end{equation}
together with Eq.~\eqref{eq:character-2}. Then Eq.~\eqref{eq:WV-linked} becomes
\begin{align}
    &\bra W_\gamma (\Gamma) V_{[U]}(x) \ket_{S^2} \nonumber \\
    &= \frac{1}{Z_{S^2}} \sum_{\alpha,\beta} N^\beta_{\alpha \gamma} \, d_\alpha\, \chi_\beta(U) \,  \rme^{-c_\alpha(\calA - \calA_\Gamma) - c_\beta \calA_\Gamma}.
\end{align}
Now taking the $\calA \to \infty$ limit, this becomes
\begin{equation}
    \bra W_\gamma(\Gamma) V_{[U]}(x) \ket
    = \chi_\gamma(U) \exp(-c_\gamma \calA_\Gamma).
\end{equation}
Comparison with Eq.~\eqref{wilson-vev} gives the first half of Eq.~\eqref{eq:WV}. 

Now consider the case where $x$ is outside of $\Gamma$. Then
\begin{align}
    &\bra W_\gamma(\Gamma) V_{[U]}(x) \ket_{S^2} \nonumber \\ 
        &= \frac{1}{Z_{S^2}} \sum_{\alpha, \beta} d_\alpha d_\beta  \exp(-c_\alpha(\calA - \calA_\Gamma) - c_\beta \calA_\Gamma) \nonumber \\
        & \quad \times \int \diff U_1 \diff U_2 \chi_\alpha (U_2 U U_2^{-1} U_1) \chi_\gamma (U_1) \chi_\beta (U_1^{-1}).
\end{align}
We evaluate the group integrals as before using formulas \eqref{eq:character-3} and \eqref{eq:character-2}, obtaining
\begin{align}
    &\bra W_\gamma(\Gamma) V_{[U]}(x) \ket_{S^2} \nonumber \\
        &= \frac{1}{Z_{S^2}} \sum_{\alpha,\beta} N^\beta_{\alpha \gamma}\, \chi_\alpha(U) \, d_\beta\, 
        \rme^{-c_\alpha(\calA - \calA_\Gamma) - c_\beta \calA_\Gamma}.
\end{align}
Taking $\calA \to \infty$, this becomes
\begin{equation}
    \bra W_\gamma(\Gamma) V_{[U]}(x) \ket
    = d_\gamma \exp (-c_\gamma \calA_\Gamma).
\end{equation}
We now get the second half of Eq.~\eqref{eq:WV}, which completes the proof.

This result \eqref{eq:WV} shows that we can measure the representation of the Wilson loop by using the topological defect operator $V_{[U]}$. 
In order to see this, let us rephrase this result in terms of canonical quantization on $S^1\times \mathbb{R}_{\mathrm{time}}$. 
Performing the canonical quantization in the temporal gauge, an eigenstate wave function is given by a Wilson loop with some irreducible representation $\alpha$ wrapping $S^1$, and let us denote it as $|\alpha\rangle$. Then, (\ref{eq:WV}) gives 
\begin{equation}
    V_{[U]}(x)|\alpha\rangle = {\chi_\alpha(U)\over d_\alpha} |\alpha\rangle. 
\end{equation}
This tells that, using the local operator $V_{[U]}$, we can construct the projection operator onto a specific state as  $|\alpha\rangle \langle \alpha|= d_\alpha \int \diff g\, \chi_{\alpha}(g^{-1})V_{[g]}(x)$.\footnote{This shows that the Hilbert space for $2$d Yang--Mills theory decomposes into distinct sectors labeled by $\alpha$.  It has been known that such a decomposition occurs with conventional $(d-1)$-form symmetry in $d$ spacetime dimensions~\cite{Hellerman:2010fv, Sharpe:2019ddn, Tanizaki:2019rbk}. In this viewpoint, we have found that non-invertible $1$-form symmetry also decomposes the Hilbert space.}

As an example, consider the case $G=\mathrm{SU}(2)$, and let us detect its adjoint test quark. 
The center symmetry, $V_{[-\mathbb{I}]}$, does not detect it because the adjoint representation has trivial $N$-ality. 
On the other hand, if we choose $U=\im \sigma_3$ for instance, then we find 
\be
\frac{\chi_{\mathrm{adj}}(\im \sigma_3)}{d_\mathrm{adj}}=-\frac{1}{3}\not=1. 
\ee
Therefore, $V_{[U]}$ can distinguish the adjoint string from the trivial one, and it allows us to explain the linear confinement of adjoint quarks very naturally. 
We also note that this operator is not invertible: By acting on the fundamental Wilson loop, we have 
\be
\frac{\chi_{\mathrm{fd}}(\im \sigma_3)}{d_\mathrm{fd}}=0, 
\ee
and thus the inverse element cannot exist. 
In this way, the non-invertible topological operators $V_{[U]}$ sucessfully explain the violation of the $N$-ality rule in $2$d Yang--Mills theory from the viewpoint of symmetry .

\section{Summary and discussion}\label{sec:summary}

In this work, we considered the question of why Casimir scaling should be exact in $2$d Yang--Mills theory. 
This has been understood as a result of dynamics, as gluons in $2$ dimensions do not propagate. 
However, it was not known if it could be understood from symmetry. 
The conventional center symmetry cannot explain why such a selection rule can exist. 
We have resolved this discrepancy by showing the existence of non-invertible $1$-form symmetry generated by the defect operator $V_{[U]}(x)$. 

This success for $2$d Yang--Mills theory is encouraging for the prospect of a similar thing happening in $3$ and $4$ dimensions. 
For $3$d Yang--Mills theory, the ground-state wave functional has been well studied numerically in Ref.~\cite{Greensite:2011pj}, based on theoretical proposals in Refs.~\cite{Greensite:2007ij, Karabali:1998yq}. 
There, it was observed that the wave functional is proportional to the Boltzmann weight of $2$d Yang--Mills theory at long distances. 
This `dimensional reduction' is supposed to be relevant to explain Casimir scaling at intermediate distances in higher dimensions, and it may give us a good hint for extending our study in $2$ dimensions to higher dimensions. 

At the same time, the $N$-ality rule should set in at large enough distances, so it seems that the non-invertible symmetry cannot be exact in $3$ or $4$ dimensions. 
While this is actually correct at finite $N$, we can still be optimistic in the $N=\infty$ theory. 
In the large-$N$ limit, the factorization theorem tells us that, for example, 
\be
\bra W_{\mathrm{adj}}(C)\ket = \left|\bra W_{\mathrm{fd}}(C)\ket \right|^2+O(N^{-2}). 
\ee
Therefore, at $N=\infty$, the adjoint confining string never breaks, and its tension must be twice as large as that of the fundamental string. 
It is an interesting problem for the future to determine whether this can be interpreted as a result of (large-$N$ emergent) non-invertible $1$-form symmetry. 

\acknowledgments
The authors thank Jeff~Greensite, Zohar~Komargodski, and Sahand~Seifnashri for a useful discussion. 
The work of Y.~T. was partially supported by JSPS KAKENHI
Grant-in-Aid for Research Activity Start-up, 20K22350. 
M.\"U. acknowledges support from U.S. Department of Energy, Office of Science, Office of Nuclear Physics under Award Number DE-FG02-03ER41260.

\bibliographystyle{utphys}
\bibliography{./QFT,./ref}
\end{document}